# Structure, Electrical and Optical Properties of ITO Thin Films and their Influence on Performance of CdS/CdTe Thin-Film Solar Cells

Moustafa Ahmed[1,*], Ahmed Bakry[1], Essam R. Shaaban[2], and Hamed Dalir[3]
[1]Department of Physics, Faculty of Science, King Abdulaziz University, 80203 Jeddah 21589, Saudi Arabia
[2]Physics Department, Faculty of Science, Al - Azhar University, P.O. 71452, Assiut, Egypt
[3]Department of Electrical and Computer Engineering, George Washington University, 20052, Washington, D.C., USA
[*]Corresponding author: mhafidh@kau.edu.sa

## Abstract

In terms of mixing graded $TiO_2$ and $SnO_2$ powders by solid-state reaction method, ITO was prepared. Using electron beam gun technology, ITO films with different thicknesses were prepared. The influence of film thickness on structure, electrical and optical properties was studied. The XRD patterns were utilized to determine the structural parameters (lattice strain and crystallite size) of ITO with different thicknesses. It is observed that the average crystallite size increases as the film thickness increases, but the lattice strain decreases. SEM shows that as the film thickness increases, the grain size of ITO increases and improves. The electrical properties of ITO films with different thicknesses were measured by the standard four-point probe method. It can be seen that as the thickness of the ITO film increases from 75 nm to 325 nm, the resistivity decreases from $29\times10^{-4}$ Ω/cm to $1.65\times10^{-4}$ Ω/cm. This means that ITO films with lower electrical properties will be more suitable for high-efficiency CdTe solar cells. Three optical layer models (adhesive layer of the substrate/B-spline layer of ITO film/surface roughness layer) are used to calculate the film thickness with high-precision ellipsometry. In the higher T(λ) and R(λ) absorption regions, the absorption coefficient is determined to calculate the optical energy gap, which increases from 3.56 eV to 3.69 eV. Finally, the effects of ITO layers of various thicknesses on the performance of CdS/CdTe solar cells are also studied. When the thickness of the ITO window layer is 325 nm, $V_{oc}$ = 0.82 V, $J_{sc}$ = 17 mA/cm$^2$, and FF = 57.4%, the highest power conversion efficiency (PCE) is 8.6%.

## 1-Introduction

High optical transparency at the same time (90%) in the visible light region and with high electricity conductivity needs to produce electronic degeneracy introduction of non-stoichiometric method in wide gap (3 eV) oxide or suitable dopants. These setting can be achieved by a variety of indium, tin, cadmium, and zinc oxides and its combination. Mixed oxides of $TiO_2$ and $SnO_2$ have attracted much consideration in the field of gas sense. Compared with pure binary oxides [1-3], their response to $H_2$ and CO is enhanced. In the field of photocatalysis, activity is obtained under ultraviolet and visible light [4-11]. In terms of gas sensing and photocatalysis, it has been initiated that adding a little amount of $SnO_2$ to $TiO_2$ can achieve the most useful outcome. The electronic structure of the material is significant to the efficient performance. Although the lattice parameters of $SnO_2$ (a = 4.594A°, c = 2.959A°) are slightly larger than that of $TiO_2$ (a = 4.737 A°, c = 3.360 A°). By reacting between $SnO_2$ and $TiO_2$ at high temperature, an alternative solid solution $Sn_xTi_{1-x}O_2$ with retained bottom red stone structure is obtained. [12,13] Above 1450 °C, a solid solution with a composition range of $0.0 < x < 1.0$ is thermodynamically stable [14]. Lower than 1450 °C, there is a nearly symmetric miscibility gap in which the solid solution decomposes spinodal into phases rich in Sn and Ti [15-16]. However, such decomposition is very slow at room temperature, and rapid quenching of the composition near the end members ($x < 0.2$, $x > 0.8$) gives long-term stable single-phase samples; [17,18] or, stable can make the composition close to room temperature under kinetic control [19]. Although the individual electronic structure of $TiO_2$ and $SnO_2$ has been studied for many years [20-22], but the electronic structure of $Sn_xTi_{1-x}O_2$ solid solution has not studied until recent [23-24]. The direct band gaps of rutile $TiO_2$ and $SnO_2$ are 3.062 and 3.596 eV, respectively [25-27]. The target of the present work divided into two folds the first is to study the effects of film thickness on the micro structural parameters (crystallize size and lattice strain) and electrical resistivity of ITO window layer. The second is calculation of film thickness with high precision in terms of spectroscopic ellipsometry.  The third is studying the film thickness effect on optical properties via measuring $T(\lambda)$ and $R(\lambda)$ of ITO window layer. The fourth is checking the impact of ITO layers with a variety of thicknesses on the performance of CdS/CdTe solar cells. The fifth is the interpretation on the change in optical parameters and the performance of CdS/CdTe solar cells in terms of microstructural parameters and electrical resistivity.

## 2. Experimental procedures

Using ball milling technology, high purity (99.9%) analytical grade powders purchased from Aldrich in stoichiometric quantities are mixed in a ball mill for about one hour. The mixed powder is then pressed into disc-shaped particles. The powder was compressed into pellets by uniaxial compression (20 MPa), and then pressed at 210 MPa. Then the pellets were sintered at 1200 °C at a heating rate of 20 °C/min in a 2-cap ambient atmosphere, and then cooled to space temperature at a rate of 20 °C/min. Such ITO particles are used as the initial material (after gridding), and the electron beam gun (Denton Vacuum DV 502 A) was used to deposit the powdered sample presence inside the quartz glass crucible at a pressure of about $10^{-6}$ Pa onto clean the glass substrate. FTM6 thickness monitor was used to monitor both the film thickness and rate of deposition. During the deposition process, the substrates were kept at temperature 100 °C and the deposition rate was adjusted at 2 nm/sec. X-ray diffraction (XRD) of powder (Philips diffraction 1710) with Cu-Kα = 1.54056 Å) was used to examine the phase clarity and crystal structure for ITO powder and films. Utilizing energy dispersive X-ray spectroscopy (EDAX) interfaced with a scanning electron microscope, SEM (JEOL JSM-6360LA, Japan) for composition analysis that confirms the relative error of the indicator element does not exceed 2.1%. The electrical properties of ITO layer films with different thicknesses were measured by a standard four-point probe method. The dual-beam spectrophotometer (UV-Vis-NIR JASCO-670) is used to measure the transmission and reflection of the film. Keithley's 2400 supply meter system is used in the solar simulator 1.5 worldwide spectrum (AM1.5G) under standard test setting to determine the current density-voltage (J-V) characteristics of solar cell equipment. More details about solar has been described in Shen et al. [28].

## 3. Results and discussion

### *3.1. Structural analysis*

Rietveld refining is a method used to characterize crystalline materials [29]. The XRD pattern of the ITO powder sample result is characterized by reflection (peak intensity) at convinced positions. The position, height and width of this reflection can be used to determine many aspects of the structure of the material. The Rietveld treatment utilizes the least squares method to treat the theoretical line contour until it matches the measured contour. Fig.1 illustrates the sample grinding of Rietveld powder to ITO. Fig. 2 shows the diffraction peaks of ITO films with different thicknesses in the XRD pattern belonging to the ITO (JCPDS data file: 39-1058-cubic), which is better oriented along the (222) plane. The main characteristics of these trends are the same, but only small differences are noticed during the peak duration. For the (222), (400), and (411) orientation planes, the diffraction angle 2θ is 30.27, 35.17, and 50.98, respectively, and suitable sharp diffraction peaks are observed. Also, Fig. 3 shows that as the thickness of the ITO film increases, the diffraction intensity of the (222) plane increases, and the increase in thickness significantly improves the crystallization efficiency of the deposited film.

The peak broadening is attributed to instrumental factors and structural factors t (that depend on lattice strain and crystal size). The crystal size (*D*) and lattice strain (*e*) are calculated by Scherrer and Wilson equations [30, 31] as follows:

$$D = \frac{0.9\lambda}{\beta\cos\theta} \tag{1}$$

$$e = \frac{\beta}{4\tan\theta} \tag{2}$$

Where $\beta$ is the width of the peak, which is equal to the difference between the width of the film and the width of standard silicon.

$$\beta = \sqrt{\beta_{obs}^2 - \beta_{std}^2}\ .$$

Fig. 4 and Fig. 5 shows the two parameters (*D* and *e*) as the function of film thickness. It is practical that the average crystallite size rises with raising the film thickness, but the lattice strain decreases. The observed micro-strain behavior may be due to the reduce in crystallite size. Similarly, the reduce in lattice strain reflects the reduce in the concentration of lattice imperfection, which may be due to the drop off in width as the thickness increases.

Fig 6 displays the SEM of three different thickness (a) d = 75 nm, (b) d = 225 nm and d = 325 nm of ITO films. It can be clearly seen from this Figures 2 that as the film thickness increases the grain size of ITO increases and improves. At the same time, the surface morphology of the ITO film becomes more compact and uniform, and as the thickness of Figure 2 increases, the crystal quality gradually improves. It should also be noted that the film is fully crystalline for higher thicknesses, which can lead to enhanced charge carrier transport and collection and further enhance device performance. Therefore, it can be concluded that the higher thickness (325 nm) of ITO film shows better crystal feature, which is also in good harmony with the earlier results of XRD. The SEM average grain size value is significantly larger than the grain size calculated by the XRD study, because the grain is composed of many grains [32].

### *3.2. Electric properties*

The electrical properties of ITO layer films with different thicknesses were measured by a standard four-point probe method. The necessary formula for sheet resistance measurement is: $Rs = 4.53 \cdot V/I$ [Ω/ sq], where: V is the voltage in volts, I is the current in amperes, and the value 4.53 is the correction constant. If the film thickness is d, the relationship between resistivity ρ (in ohm cm) and Rs is given by $Rs = \rho/d$ [33].

The relationship between the resistivity of the ITO film and the film thickness is shown in Fig. 7. It can be seen from Fig. 7 that as the ITO film thickness increases from 75 nm to 325 nm, the resistivity decreases from $29 \times 10^{-4}$Ω/cm to$1.65 \times 10^{-4}$ Ω/cm, respectively. The decrease in resistivity leads to a relatively high charge carrier density, which in turn causes mobility, which may be attributed to the relatively high crystal quality and larger grain size and excess cations (In or Sn), all of which both lead to an increase in charge carriers and a decrease in grain boundary scattering. This means that ITO films with lower electrical properties will be more appropriate for high-efficiency CdTe solar cells.

### *3.3. Optical properties*

For ITO films of different thicknesses, a dual beam spectrophotometer can be used to obtain the measured values of optical transmittance (*T*) and reflectance (*R*) relative to wavelength. Fig. 8 shows T(λ) and R(λ) with respect to the wavelength λ. The transmittance is found to be decreased with increasing the film thickness particulary in NIR region. In the near-infrared region, due to the large number of free electrons in the film, the interaction between free electrons and incident light occurs. This interaction may cause the polarization of the light in the film, which causes a remarkable decreasing in the transmission spectra, thereby affecting the dielectric constant. Transmittance is highly dependent on thickness of prepared thin films. But the reflectance spectra of the same set of films. It shows that for wavelengths above 1900 nm, the reflectivity will increase slightly. However, this increase in reflectance does not coincide with the decrease in transmittance in the same area. Therefore, as the thickness of the near-infrared region increases, the decrease in transmittance is attributed to free carrier absorption, which is common in all transparent conductors with high carrier concentration [34].

The film thickness is calculated by the spectroscopic ellipsometry parameters (ψ and Δ), which are measured in the wavelength range of 380-880 nm using a rotating compensator instrument (J.A. Woollam, M-2000). All the details of this method have been clearly seen in references [35-37]. The data was acquired at a 70° incident angle. According to the developed WVASE32 program, as shown in Figure 9, three optical layer models ((Cauchy layer of substrate /B-spline layer of ITO film/surface roughness layer.)) are used to determine the film thickness with high accuracy.

In a higher absorption region for both *T*(λ) and *R*(λ), the absorption coefficient, α can be derived from the following expression[38, 39]:

$$\alpha = \frac{1}{d}\ \ln\left[\frac{(1-R)^2 + \left[(1-R)^4 + 4R^2T^2\right]^{1/2}}{2T}\right] \tag{6}$$

where $d$ is the thickness of the film. Fig. 10 shows the reliance of α ($h$ν) on photon energy, $h$ν as a function of film thickness. Pure semiconducting compounds are known to have a sharp absorption edge [38, 39]. The edge of absorption was sharper and moved to higher wavelengths, as the film thickness rose from 75 nm to 325 nm. It is known that the α value is described in the higher neighborhood of the fundamental absorption edge (higher $10^4$ $cm^{-1}$), for allowed direct transition from valance band to conduction band. But the value of α at energy rang extended from 1 to 2 eV represent a transparent visible region. The energy gap values can be calculated using Tauc relation [40] as follows

$$\alpha(h\upsilon) = \frac{K\,(h\upsilon - E_g^{opt})^m}{h\upsilon} \tag{7}$$

Where K is a parameter independent of $h$ν for the individual transitions [35], $E_g^{opt}$ is the optical energy gap, and $m$ is a number that identify the type of the transition. Various authors [41-43] recommended different $m$ values, such that $m$ = 2 for the majority amorphous semiconductors (indirect transition) and m = 1/2 for most crystalline semiconductors (direct transition). At different annealing temperature of Cu doped CdTe film, the direct transition is valid. Fig 11 illustrates the best fit of $(\alpha h\nu)^2$ vs. $h\nu$ for varies thickness. The direct optical band gap $E_g^{opt}$ can be taken as the intercept of $(\alpha.\ h\nu)^2$ against. ($h\nu$) at $(\alpha\ h\nu)^2 = 0$. The $E_g^{opt}$ resulting for each film is shown in Fig. 10. The $E_g^{opt}$ increase as the film thickness increases. The value of the optical band gap $E_g^{opt}$ was found to vary from 3.56 in the film for d = 75 nm to 3.69 eV for d = 325 nm. This increase was may attributed to reasons the first is a reduction in the resistivity of the films, implying an enhancement of the carrier density as in Fig. 7, and this change is well known as the Burstein–Moss shift [44] and the second is due to the increasing in crystallites size i.e increasing of the crystallinity of the film with increasing the thickness.

### *3.4. Impact of thickness of ITO window layer in the performance CdS/CdTe solar cells*

The schematic diagram of the configuration of the CdS/CdTe solar cell is shown in Fig. 11. Thus, the device configuration is a simple n-p heterojunction with an ohmic contact at the p-CdTe/metal interface. In order to show the influence of ITO layer thickness on the performance of CdS/CdTe solar cells, CdS/CdTe cells based on varies ITO window layer thicknesses were equipped. Figure 12 (a, b and c) shows the current density-voltage (J-V) characteristics of the CdS/CdTe cells (with the ITO layer thickness is 75nm, 225 nm and 325 nm) and the corresponding photovoltaic parameters are shown in these figures. Fig. 12 (c) was clarified that the highest power efficiency (PE) is 8.6 % with $V_{oc}$ = 0.82 V, $J_{sc}$ = 17 mA/ $cm^2$, FF = 57.4% at the ITO layer thickness is 325 nm. We think that this development in the device performance may be associated to the high transmittance value and lower reflectivity under this ITO layer thickness. In addition, the $J_{sc}$ of the cells with the ITO window layer of 75 nm and 225 nm is lower than the Jsc of the cell with the ITO window layer of 325 nm. The cause for the elevated in Voc and Jsc based on the 325 nm ITO layer may be due to the development of photoelectric characteristics, because due to the decrease in reflectivity, a large number of photons are absorbed into the CdS/CdTe solar cell, resulting in more absorption in the absorption layer. Photo-generated carriers and further enhanced Voc and Jsc. Additionally, the main reason for the highest $V_{oc}$ and $J_{sc}$ may be attributed to the first is an increasing in grain size thus improvement of cystallinity with the larger thickness and the second is due to smaller resistivity. The lower resistivity of 325 nm ITO layer can take part to the development of the FF of CdS/CdTe solar cells. Combining with the above analysis, it can be concluded that 325 nm ITO window layer is beneficial for the improvement of PEC for CdS/CdTe solar cells. Based on the above study, it can be accomplished that the 325 nm ITO window layer is useful to the development of the PE of CdS / CdTe solar cells.

Beyond solar cells specifically, such material characterization for light-matter interactions bears a high-relevance for photo conversion applications such as solar cells. In fact, recent work on window-integrated PC concentrators [45,46] has shown to deliver both renewable power generation whilst allowing for spectral-tuning of the optical structure to allow for natural room lighting and reduced air conditioning load on a dwelling by designing photonic structures (e.g., high contrast gratings) to reflect the infrared spectrum (reducing heat load) yet transmitting the visible portion. Indeed, in this study we found that interplay between material quality to solar cell performance can be extended, in principle, to high absorption atomically flat van-der Waals materials such as those belonging to the class of transition metal dichalcogenides (TMD) for light emission [47] and light detection [48, 49]. In fact, recent work points to some interesting scaling laws in photodetectors combining such emerging highly-absorptive materials with scaling-length-theory known from transistor short-channel devices, enabling a new class of high-gain-bandwidth product photodetectors [50].

Furthermore, controlling the ITO material properties during the deposition process is critical given plurality of control parameters of ITO such as activating the Sb dopants via annealing or adding/subtracting oxygen [51, 52]; both impact the complex optical index, and hence the transport properties such as the resistivity and mobility. Our material property findings for ITO in this study are indeed far-ranging beyond photo-absorption but can further be used to electro-optic devices as well [53-56]. For instance, the optical index can be tuned electrostatically via chancing the free carrier concentration leading to electro-optic modulators, both for electro-absorption [57-59], but also used in interferometric schemes [60-62] to demonstrate high-performance modulators.This tunability of ITO thin films can be uniquely utilized also for 2x2 switches in integrated photonics [63], as electro-optic nonlinear activation function in photonic neural networks [64], or as a phase-shifter for optical phase array applications for LiDAR systems [65]. Interestingly, the unique properties of ITO have recently inspired applications for photonic and nano-optic analog computing schemes such as for solving partial differential equations [66,67].

## *4. Conclusions*

In this work, the structural, morphological, electrical, and optical of ITO films with various thicknesses were investigated. Besides, the effect of the different ITO layer thicknesses on the properties of CdS/CdTe solar cells were characterized. The film thickness was calculated using spectroscopic ellipsometry with high precision using three optical layer models (Cauchy layer of substrate /B-spline layer of ITO film/surface roughness layer). The results indicated that the crystallites size identified via XRD pattern and grain size determined by SEM were increased with raising the ITO film thickness. The resistivity decreased sharply from 29 x $10^{-4}\Omega$/cm to1.65 x $10^{-4}$ $\Omega$/cm with increasing ITO film thickness from 75 nm to 325 nm. The transmittance spectra found to be decreased with increasing the film thickness particularly in NIR region due to the large number of free electrons in the film, the interaction between free electrons and incident light occurs. The optical band gap was calculated in terms of transmittance and reflection spectrum in the high region of the absorption. In terms of Tauc's relationship, the possible transition of ITO films found to be allowed direct transition with increasing of the band gap from 3.56 to 3.69 eV with an increase of the film thickness from 75 nm to 325 nm. The impacts of ITO layers with various thicknesses on the performance of CdS/CdTe solar cells were also studied. The highest power conversion efficiency (PCE) is 8.6 % with $V_{oc}$ = 0.82 V, $J_{sc}$ = 17 mA/ $cm^2$, FF = 57.4% when the ITO window layer thickness is 325 nm.

**Acknowledgments**

This project was funded by the Deanship of Scientific Research (DSR) at King Abdulaziz University, Jeddah, under grant no. (**RG-38-130-41**). The authors, therefore, acknowledge with thanks DSR technical and financial support.

## References

[1] K. Zakrzewska, *Thin Solid Films*, 391, 229–238 (2001).

[2] W. P Tai, J. H. Oh, *Sens. Actuators*, B 85,154–157 (2002).

[3] M.C. Carotta, S. gherardi, V. Guidi, C. Malagù, G. Martinelli, B. Vendemiati, M. Sacerdoti, G. Ghiotti, S. Morandi, A. Bismuto, P. Maddalena, A. Setaro, *Sens. and Actuators B.* 130, 38–45 (2008).

[6] Z. Q. Cai, Q. H. Shen, J. W.Gao, H. J.Yang, *Inorg. Mater*. 22, 733–736 (2007).

[7] F. Fresno, C. Guillard, J. M. Coronado, Chovelon, J. M. Tudela, D. J. Soria, J. M. Herrmann, *J. Photochem. Photobiol. A.* 173,13–20 (2005).

[8] F. Fresno, J. A. Coronado, D. Tudela, J. Soria, *Appl. Catal. B.,* 55,159–167 (2005).

[9] L. Q. Jing, H. G. Fu, D. J. Wang, X. Wei, J. Z. Sun, *Acta Phys.-Chim. Sin.* 21, 378–382 (2005).

[10] Y. A. Cao, W. S. Yang, W. F. Zhang, G. Z. Liu, P. L. Yue, *New J. Chem*. 28, 218–222 (2004).

[11] C. F. Lin, C. H. Wu, Z. N. Onn, *J. Hazard. Mater*. 154,1033–1039 (2008).

[12] F. Sayilkan, M. Asiltuerk, P. Tatar, N. Kiraz, S. Sener, E. Arpac, H. Sayilkan, *Mater. Res. Bull*. 43,127–134 (2008).

[13] M. Emam-Ismail, M. El-Hagary, E. R. Shaaban, SH Moustafa, GMA Gad Ceramics International 45 (7) 8380-8387 (2020).

[14] A. F. AL Naim, A. Solieman, E. R. Shaaban, *J. Mater. Sci.: Materials in Electronics* 31 (4) 3613-3621 (2020).

[15] S. Nambu, M. Oiji, *J. Am. Ceram. Soc*. 74, 1910–1915 (1991).

[16] L. B. Kong, J. Ma, H. Huang, *J. Alloys Compd*. 336, 315–319 (2002).

[18] A. Qasem, E. R. Shaaban, M. Y. Hassaan, M. G. Moustafa, M. A. S. Hammam *J. Electron. Mater*. 49 (10), 5750-5761 (2020).

[19] S. K. Kulshreshtha, R. Sasikala, V. Sudarsan, *J. Mater. Chem*. 11, 930–935 (2001).

[20] A. M. Abdelraheem, M. I. Abd-Elrahman, M. Mohamed, N. M. A. Hadia, E. R. Shaaban, *J. Electron. Mater*. 49, 1944-1956 (2020).

[21] A. M. Mazzone, *Philos. Mag. Lett.* 84, 275–282 (2004).

[22] C. M. Freeman, C. R. A. Catlow, *J. Solid State Chem*. 85, 65–75 (1990).

[23] D. F.Cox, T. B. Fryberger, S. Semancik, *Phys. Rev. B*. 38, 2072–2083 (1988).

[24] M. Batzill, U. Diebold, *Prog. Surf. Sci*. 79, 47–154 (2005).

[25] K. M. Glassford, J. R. Chelikowsky, *Phys. Rev. B*. 46, 1284– 298 (1992).

[26] F. R. Sensato, R. Custodio, E. Longo, A. Beltran, J. Andres, *Catal. Today* 85, 145–152 (2003).

[27] J. R. Sambrano, G. F. Nobrega, C. A. Taft, J. Andres, A. Beltran, *Surf. Sci.* 580, 71–79 (2005).

[28] P.-S. Shen, C-M. Tseng, T-C. Kuo, C-K. Shih, M-H. Li, P. Chen, *Solar Energy* 120, 345–356 (2015).

[29] H. M. Rietveld, *J. Appl. Cryst*. 2, 65–71 (1969).

[30] E. R. Shaaban, I. Kansal, S. H. Mohamed, J. M. F. Ferreira, *Physica B: Condensed Matter*. 404, 3571 (2009).

[31] J. Zhang, L. Feng, W. Cai, J. Zheng, Y. Cai, B. Li, L. Wu, Y. Shao, *Thin Solid Films* 414, 113–118 (2002).

[32] A. Goktas, F. Aslan, I. H. Mutlu, *J. Mater. Sci.: Mater. Electron*. 23, 605–611 (2012).

[33] R. Jaeger, *Introduction to Microelectronic Fabrication (2nd ed.),* New Jersey: Prentice Hall. pp. 81–88. ISBN 0-201-44494-1 (2002).

[34] E. R. Shaaban, *J. Alloys Compd*. 563, 274 (2013).

[35] D. Pereda Cubian, M. Haddad, R. Andre, R. Frey, G. Roosen, J. Arce, C. Diego, L. Flytzains, *Physical Review B*. 67, 45308 (2003).

[36] M. Emam-Ismail, E.R. Shaaban, M. El-Hagary, I. Shaltout, *Philosophical Magazine* 90, 3499 (2010).

[37] M. El-Hagary, M. Emam-Ismail, E.R. Shaaban, I. Shaltout, *J. Alloys Compd*. 485, 519 (2009).

[38] M. Mohamed, A. M. Abdelraheem, M. I. AbdElrahman, N. M. A. Hadia, E. R. Shaaban, *Applied Physics A*. 125:483 (2019).

[39] M. Mohamed, E. Shaaban, M. N. Abd-el Salam, A. Abdel-Latief, S.A. Mahmoud, M. Abdel-Rahim, Optik, 178, 1302-1312 (2019).

[40] C. Huang, H. Weng, Y. Jiang, H. Ueng, *Vacuum* 83, 313 (2009).

[41] E. Bacaksiz, S. Aksu, N. Ozer, M. Tomakin, A. Ozcelik, *Appl. Surf. Sci.* 256, 1566 (2009).

[42] J. Tauc, R. Grigorovici, A. Vancu, *Physica Status Solidi (b)* 15.2, 627-637 (1966).

[43] T. Mahalingam, V. Dhanasekaran, R. Chandramohan, J.K. Rhee, *Journal of Materials Science* 47, 1950 (2012).

[44] J. H. Kim, B. D. Ahn, C. H. Lee, K. A. Jeon, H. S. Kang, G. H. Kim and S. Y. Lee, *Thin Solid Films* 515, 3580 (2007).

[45] A. Elikkottil, M. Tahersima, S. Gu, V. J. Sorger, B. Pesala, *Sci. Rep*. 9, 11723 (2019).

[46] S. Gupta, M. H. Tahersima, V. J. Sorger, B. Pesala, *Opt. Express* 28, 15 (2020).

[47] M. H. Tahersima, M. D. Birowosuto, Z. Ma, W. C. Coley, M. D. Valentin, S. N. Alvillar, I-H. Lu, Y. Zhou, I. Sarpkaya, A. Martinez, I. Liao, B. N. Davis, J. Martinez, D. Martinez, A. Guan, A. E. Nguyen, K. Liu, C. Soci, E. Reed, L. Bartels, V. J. Sorger, *ACS Photonics* 4, 1713-1721 (2017).

[48] M. Tahersima, V. J. Sorger, *Nanotechnology* 26, 344005 (2015).

[49] R. Maiti, C. Patil, T. Xie, J. Ghasemi, M.A.S.R. Saadi, R. Amin, M. Miscuglio, D. Van Thourhout, S. D. Solares, T. Low, R. Agarwal , S. Bank, V. J. Sorger, *Nat. Photonics* 14, 578-484 (2020).

[50] V. J. Sorger, R. Maiti, *Opt. Mater. Express*, 10, 2192-2200 (2020).

[51] Y. Gui, M. Miscuglio, Z. Ma, M. T. Tahersima, S. Sun, R. Amin, H. Dalir, V. J. Sorger, *Nature Scientific Reports* 9, 1-10 (2019).

[52] Z. Ma, Z. Li, K. Liu, C. Ye, V. J. Sorger, Nanophotonics 4, 198-213 (2015).

[53] R. Amin, J. B. Khurgin, V. J. Sorger , *Opt. Express* 26 (11), 15445-15470 (2018).

[54] V. J. Sorger, R. Amin, J. B. Khurgin, Z. Ma, H. Dalir, S. Khan, *Journal of Optics* 20, 014012 (2018).

[55] C. Ye, S. Khan, Z.R. Li, E. Simsek, V. J. Sorger , *IEEE Selected Topics in Quantum Electronics* 4, 20 (2014).

[56] S. K. Pickus, S. Khan, C. Ye, Z. Li, V. J. Sorger, *IEEE Photonics Society* 27, 6 (2013).

[57] V. J. Sorger, D. Kimura, R.-M. Ma, X. Zhang, Nanophotonics 1,17-22 (2020).

[58] C. Huang, R. J. Lamond, S. K. Pickus, Z. R. Li, V. J. Sorger, *IEEE Photonics Journal* 5 (4), pp. 2202411-2202411 (2013).

[59] M. H. Tahersima, Z. Ma, Y. Gui, S. Sun, H. Wang, R. Amin, H. Dalir, R. Chen, M. Miscuglio, V. J. Sorger, *Nanophotonics* 8, 9 (2019).

[60] R. Amin, R. Maiti, C. Carfano, Z. Ma, M. H. Tahersima, Y. Lilach, D. Ratnayake, H. Dalir, V. J. Sorger, *Apl. Photonics* 3, 12 (2018).

[61] R. Amin, R. Maiti, Y. Gui, C. Suer, M. Miscuglio, E. Heidari, R. T. Chen, H. Dalir, V. J. Sorger, *Optica* 7, 3 (2020).

[62] R. Amin, R. Maiti, Y. Gui, C. Suer, M. Miscuglio, E. Heidari, J. B. Khurgin, H. Dalir, V. J. Sorger, arXiv: 2007:15457 (2020).

[63] C. Ye, K. Liu, R. Soref, V. J. Sorger, Nanophotonics 4(3) 261-268 (2015).

[64] R. Amin, J. George, S. Sun, T. Ferreira de Lima, A. N. Tait, J. B. Khurgin, M. Miscuglio, B. J. Shastri, P. R. Prucnal, T. El. Ghazawi, V. J. Sorger, *APL Materials* 7, 081112 (2019).

[65] M. Miscuglio, X. Ma, T. El-Ghazawi, T. Itoh, A. Alu, V. J. Sorger, arXiv preprint: 2007.05380 (2020).

[66] R. Amin, R. Maiti, J. K. George, X. Ma, Z. Ma, H. Dalir, M. Miscuglio, V. J. Sorger, *J. Light. Technol*. 38(2) 282-290 (2019).

[67] H. Dalir, F. Mokhtari-Koushyar, I. Zand, E. Heidari, X. Xu, Z. Pan, S. Sun, R. Amin, V. J. Sorger, R. T. Chen “Atto-Joule, high-speed, low-loss plasmonic modulator based on adiabatic coupled waveguides” Nanophotonics, 7(5), 859-864 (2018).

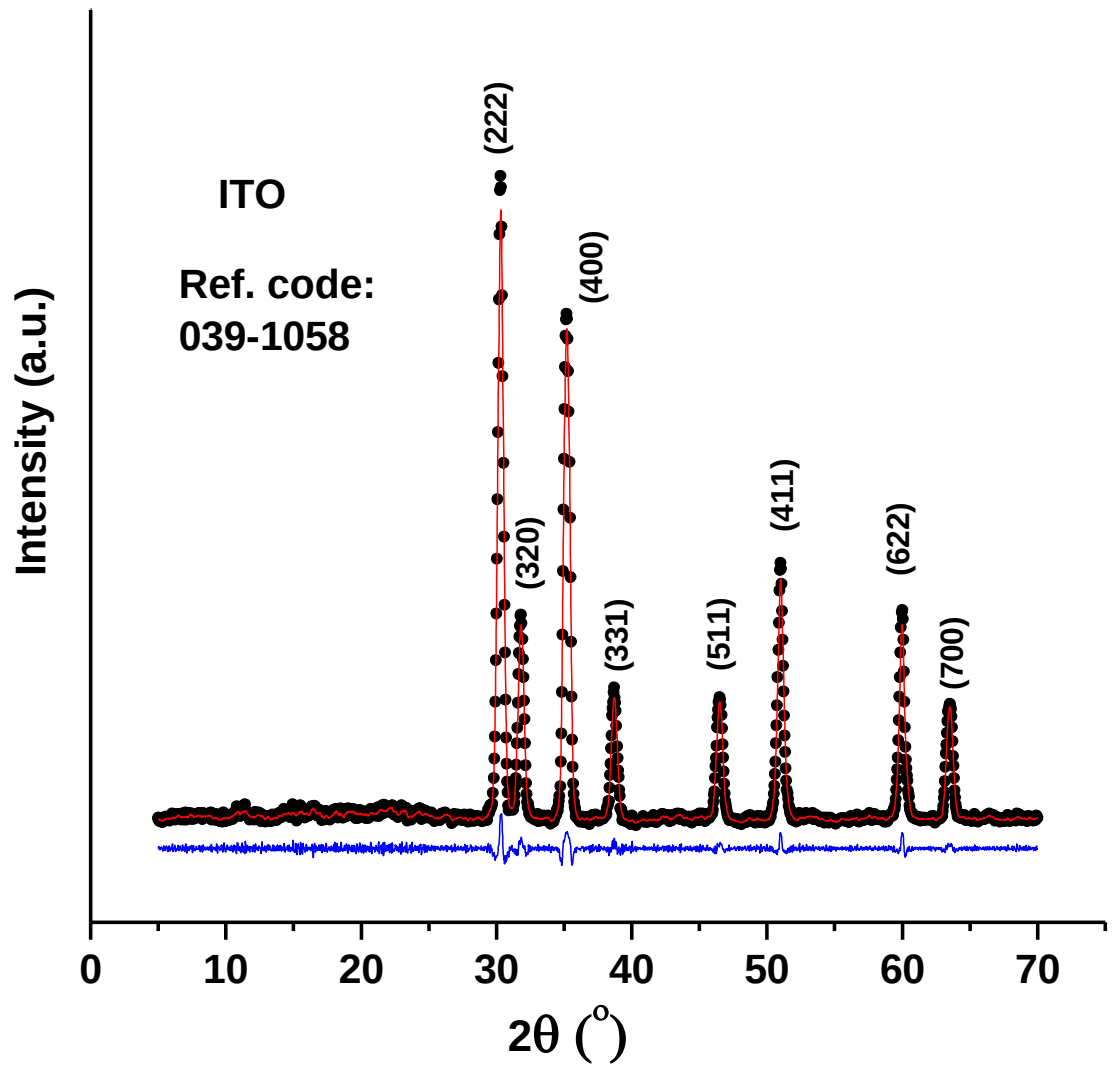

**Fig. 1:** XRD patterns and Rietveld refinement of polycrystalline ITO powder sample.

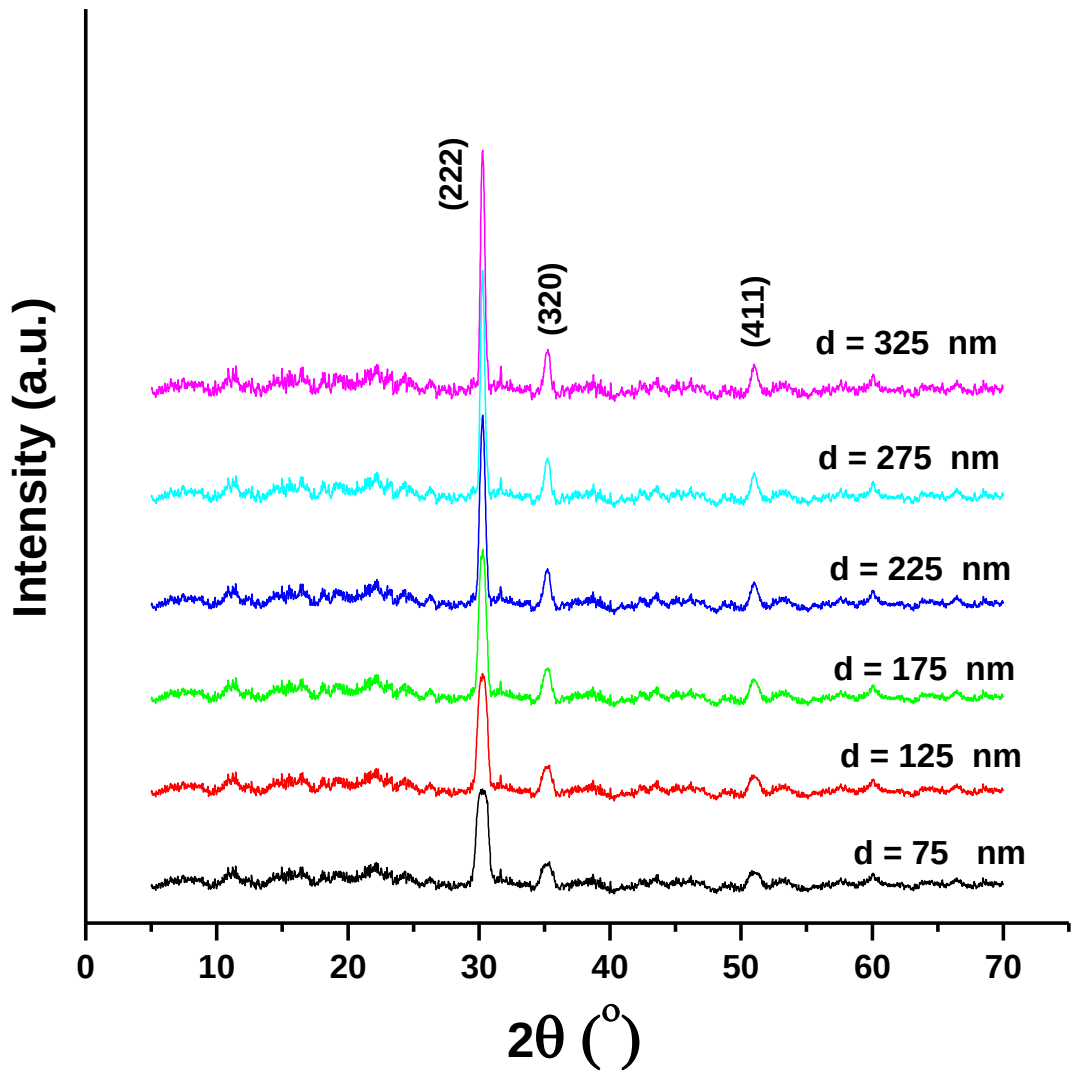

**Fig. 2:** XRD patterns of the ITO films as a function of different thickness

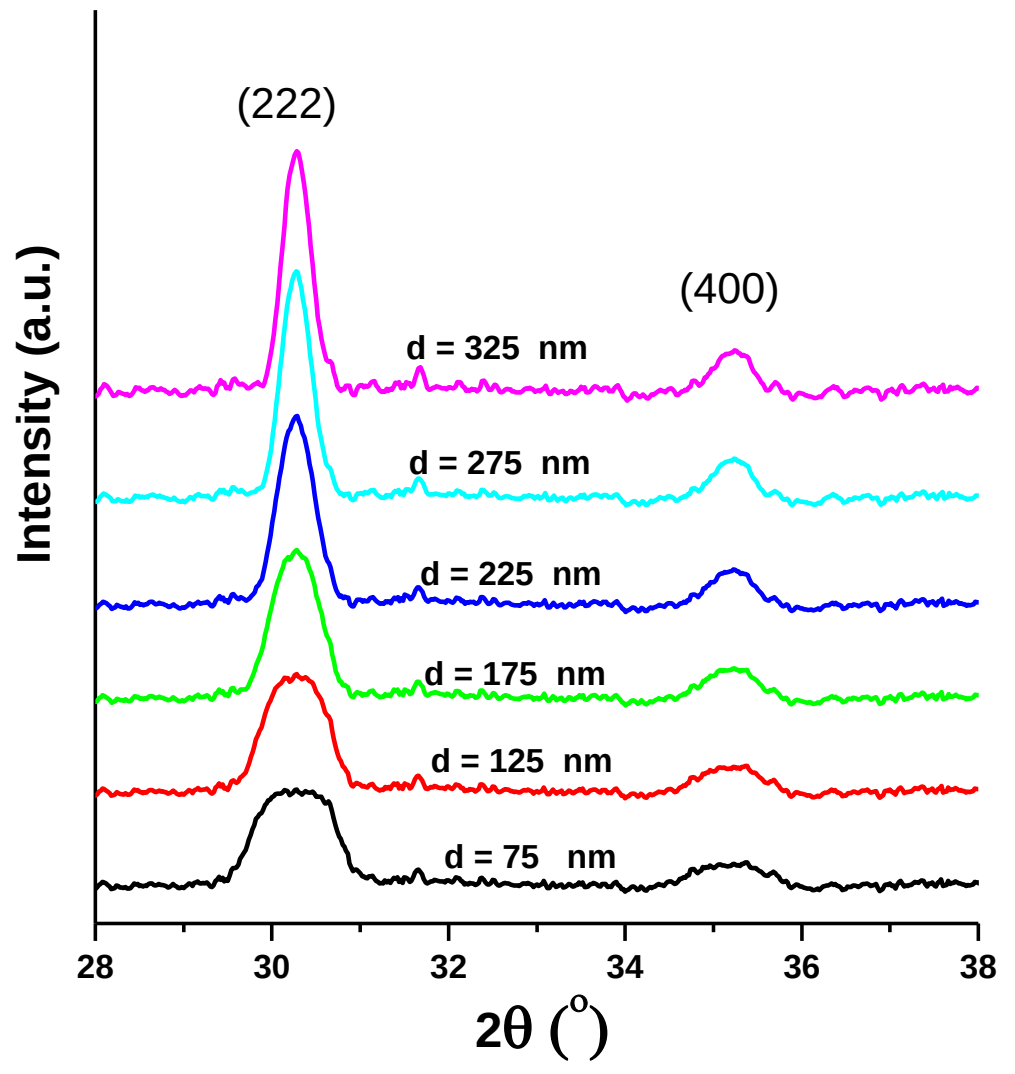

**Fig. 3:** Amplification of (222) and (400) scattering peaks of ITO a function of different thickness

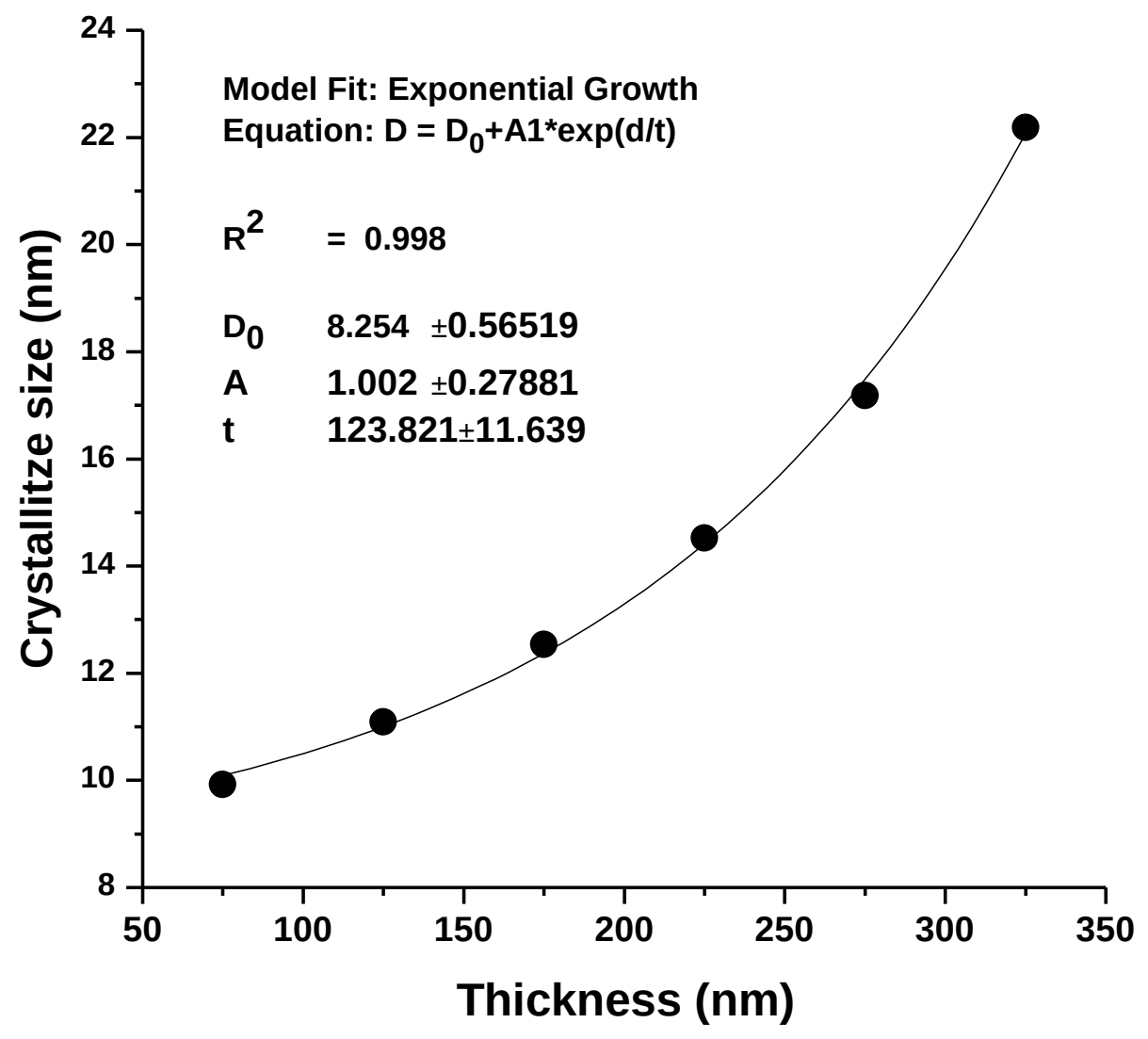

**Fig. 4:** Crystallize size versus thickness of ITO films

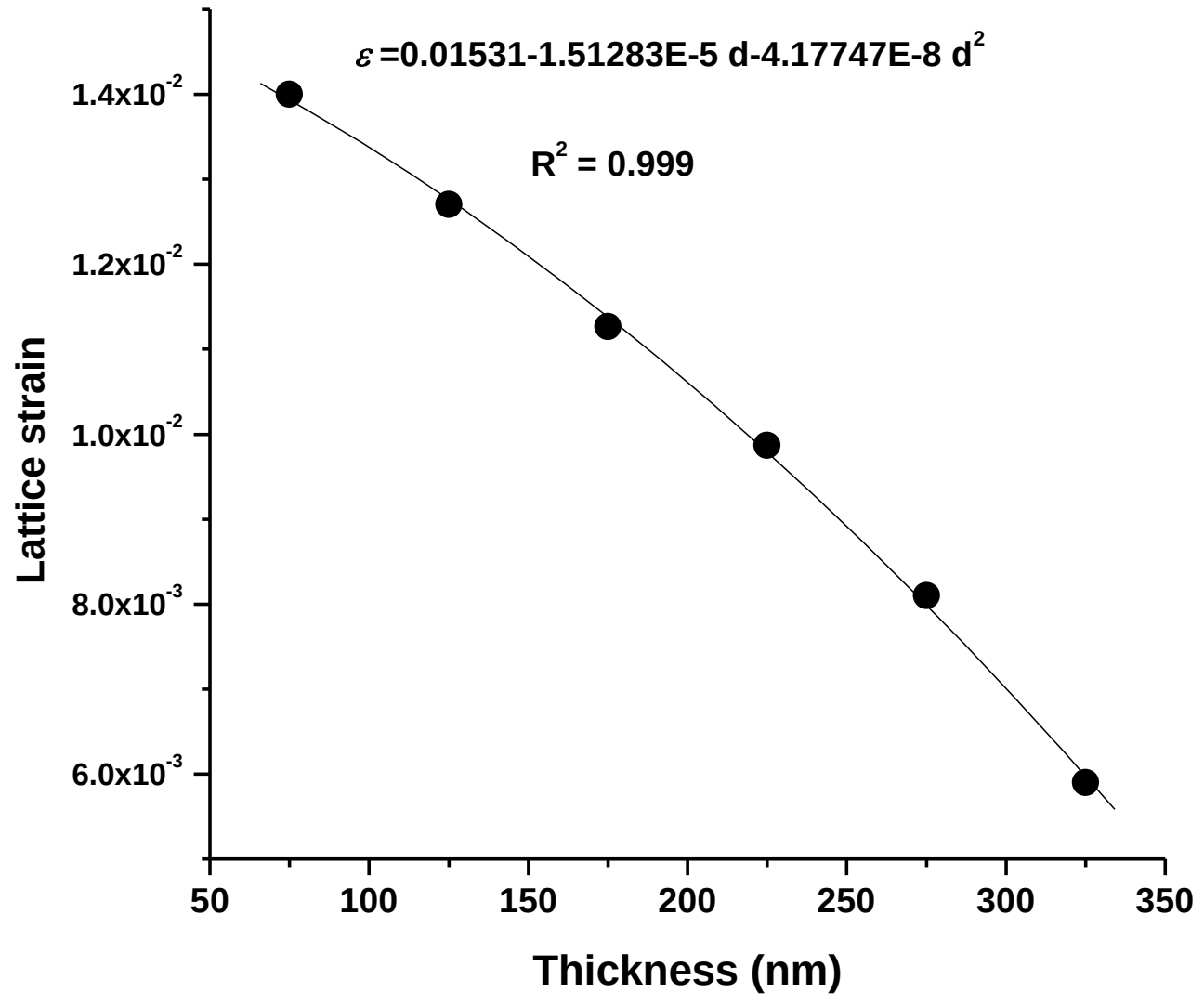

**Fig. 5:** Lattice strain versus thickness of ITO films

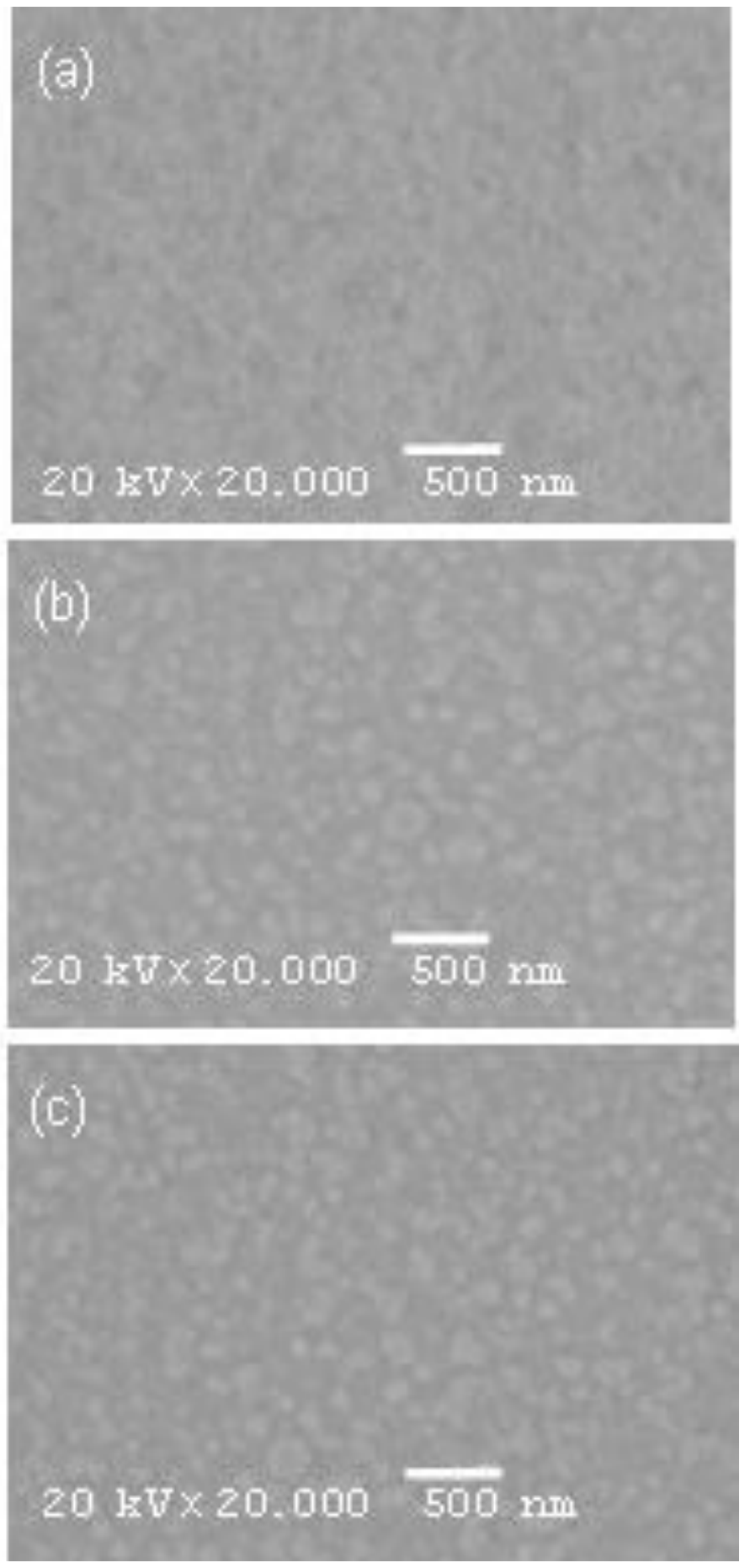

**Fig. 6:** SEM of three different thickness (a) d = 75 nm, (b) d = 225 nm and d = 325 nm.

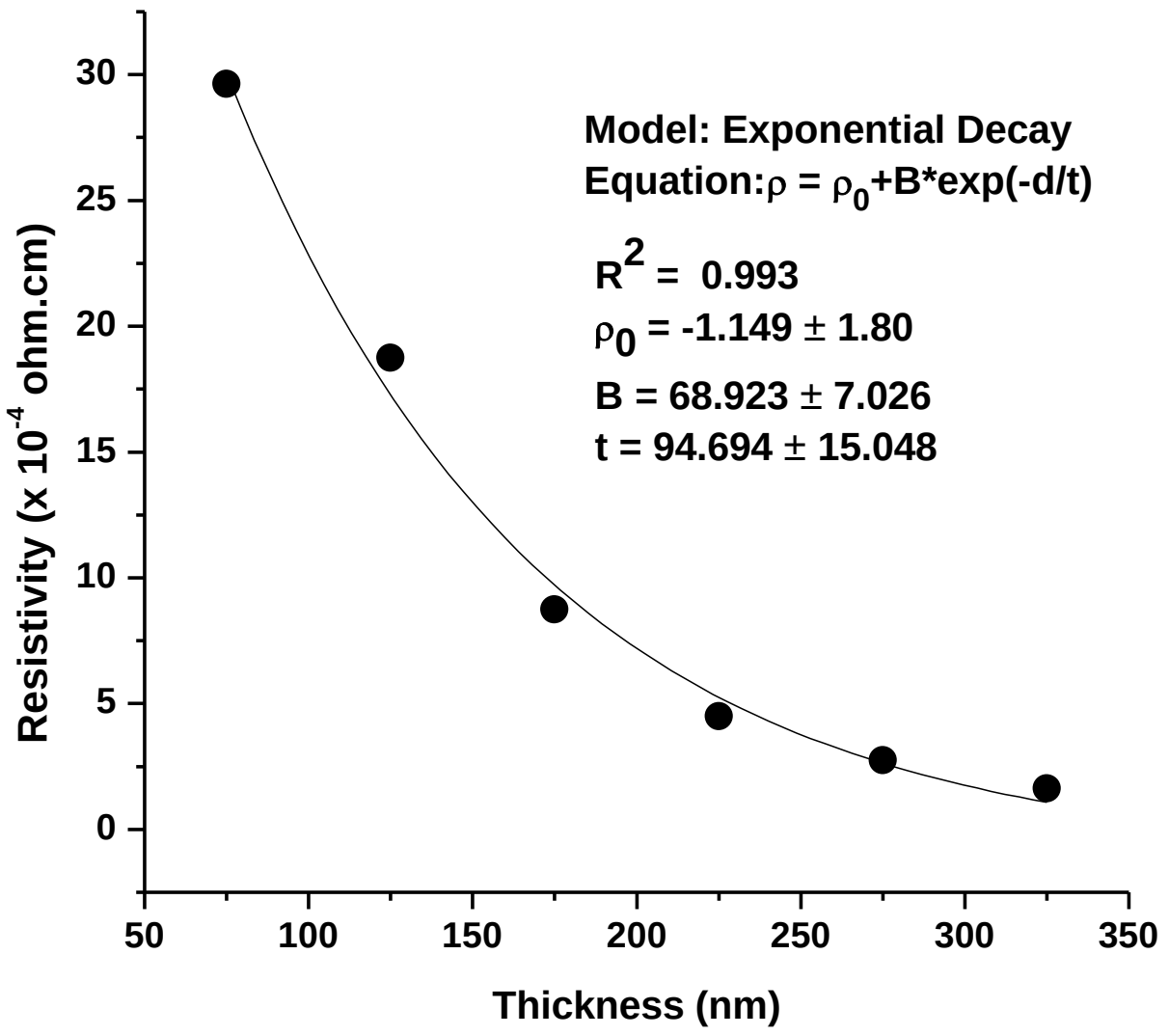

**Fig. 7:** Electric resistivity versus thickness of ITO thin films**.**

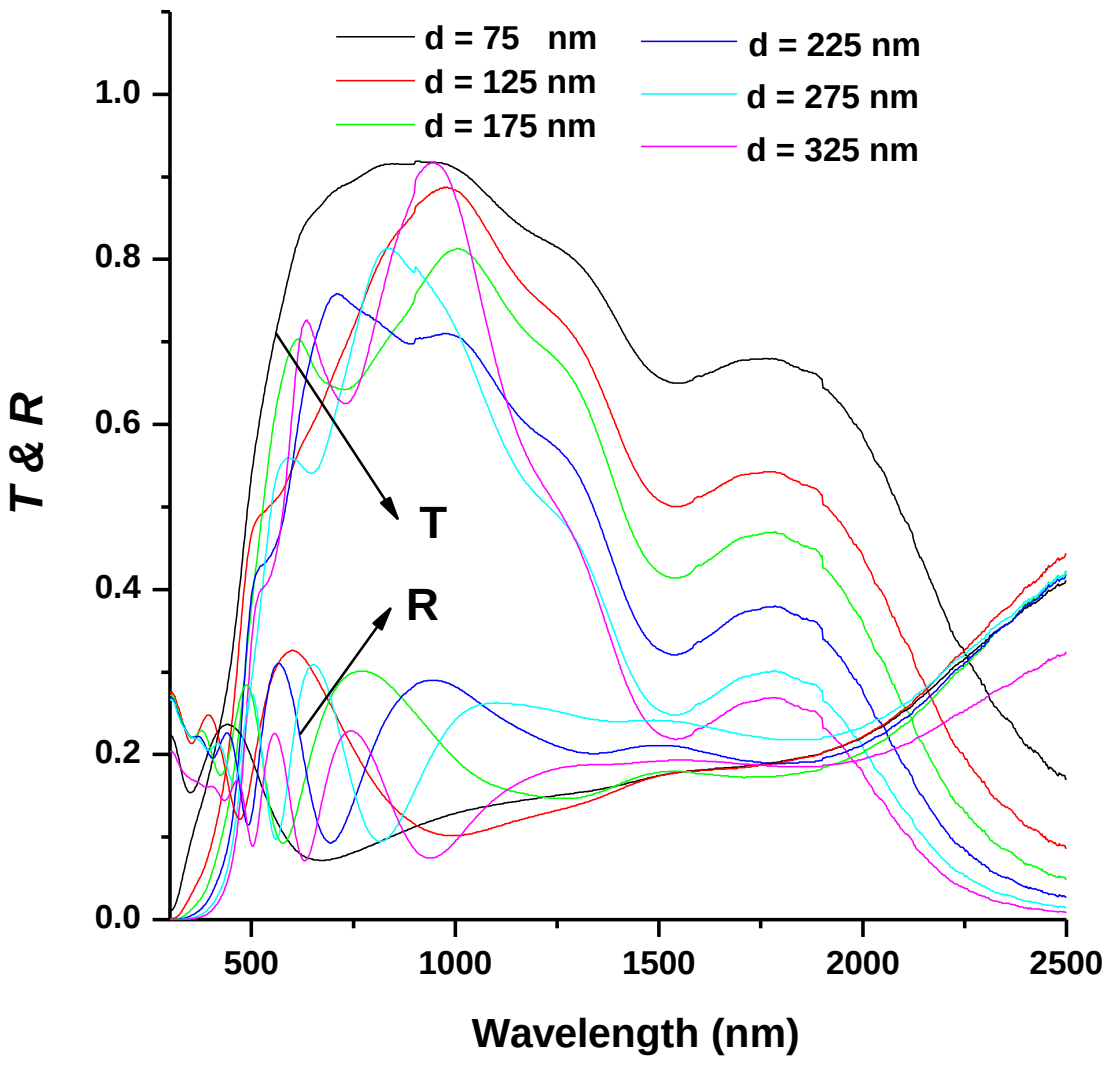

**Fig. 8:** Experimental transmission and reflection of different thickness of ITO thin films.

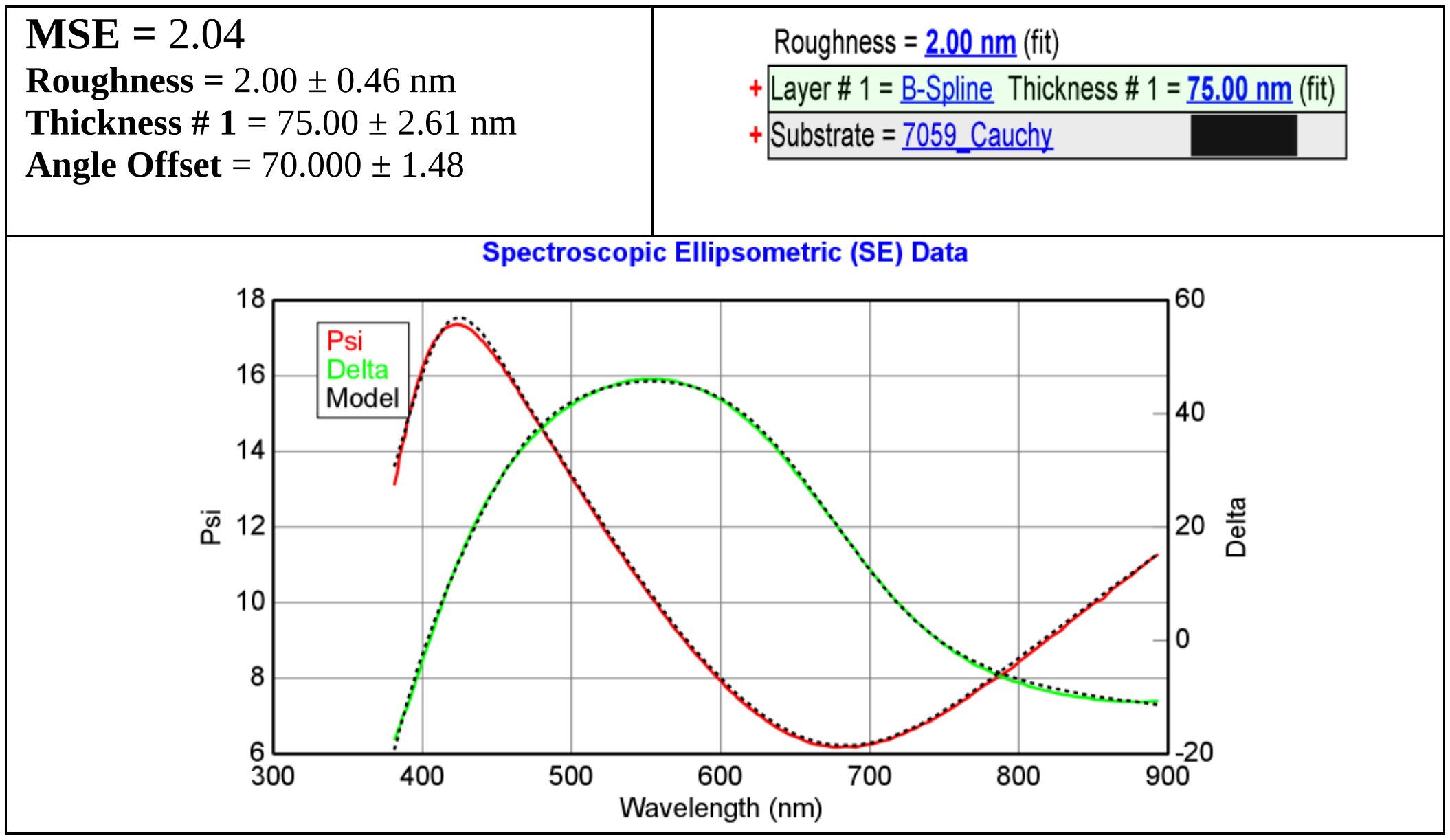

**Fig. 9**: Experimental and modeled ellipsometric optical parameters Psi and Delta for calculating the film thickness of ITO.

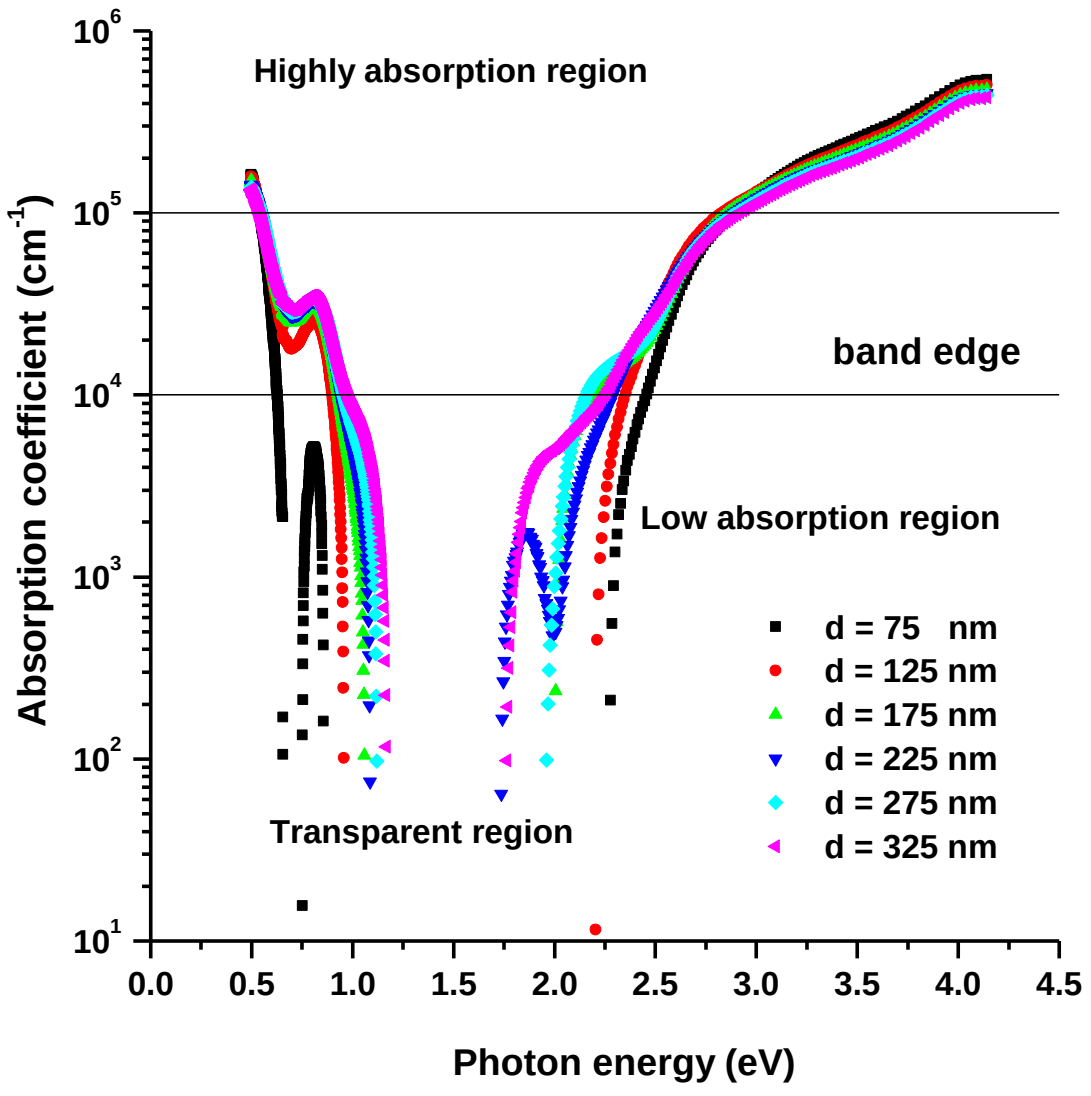

**Fig. 10:** Absorption coefficient versus photon energy of different thickness of ITO thin films.

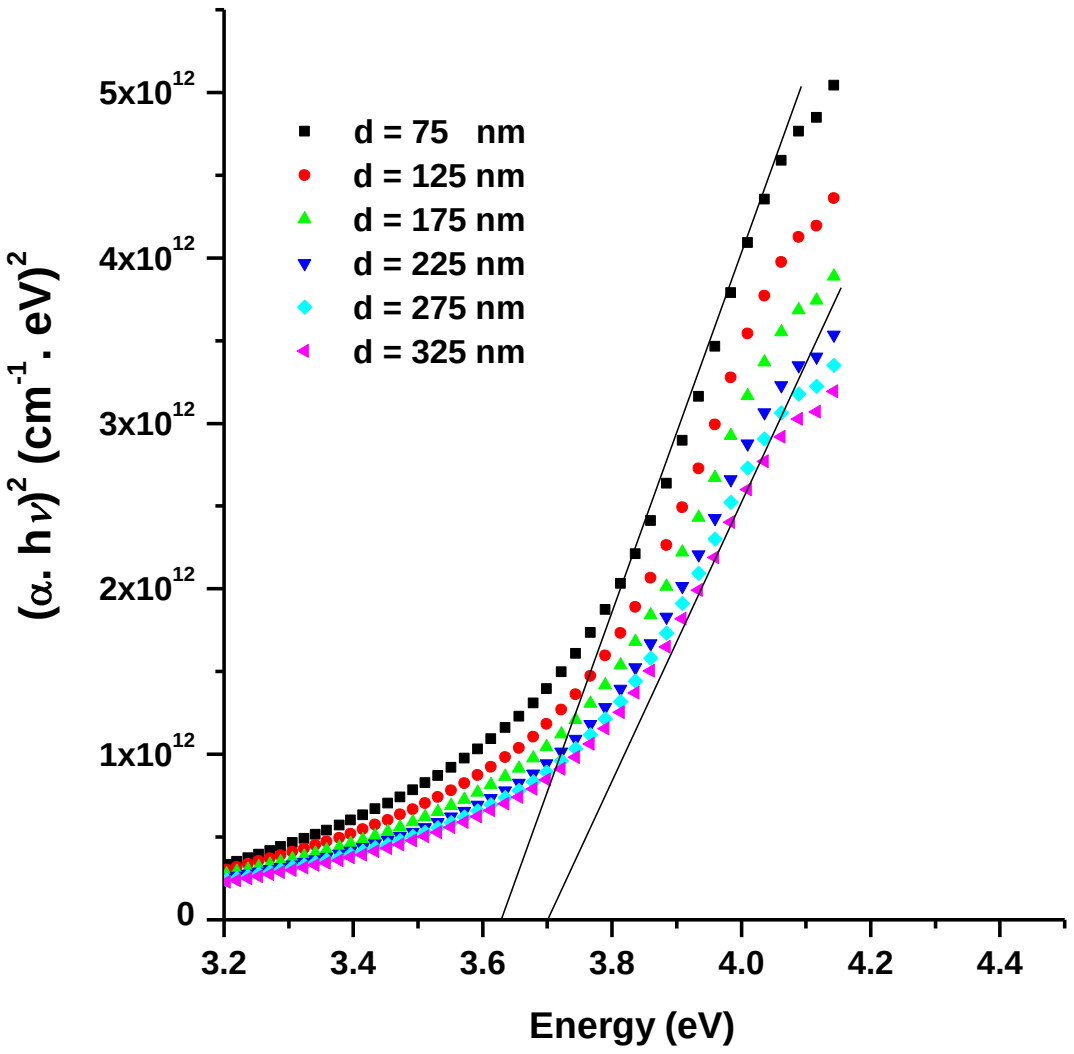

**Fig. 11:** $(\alpha h\nu)^2$ vs. photon energy $h\nu$ for of different thickness of ITO thin films.

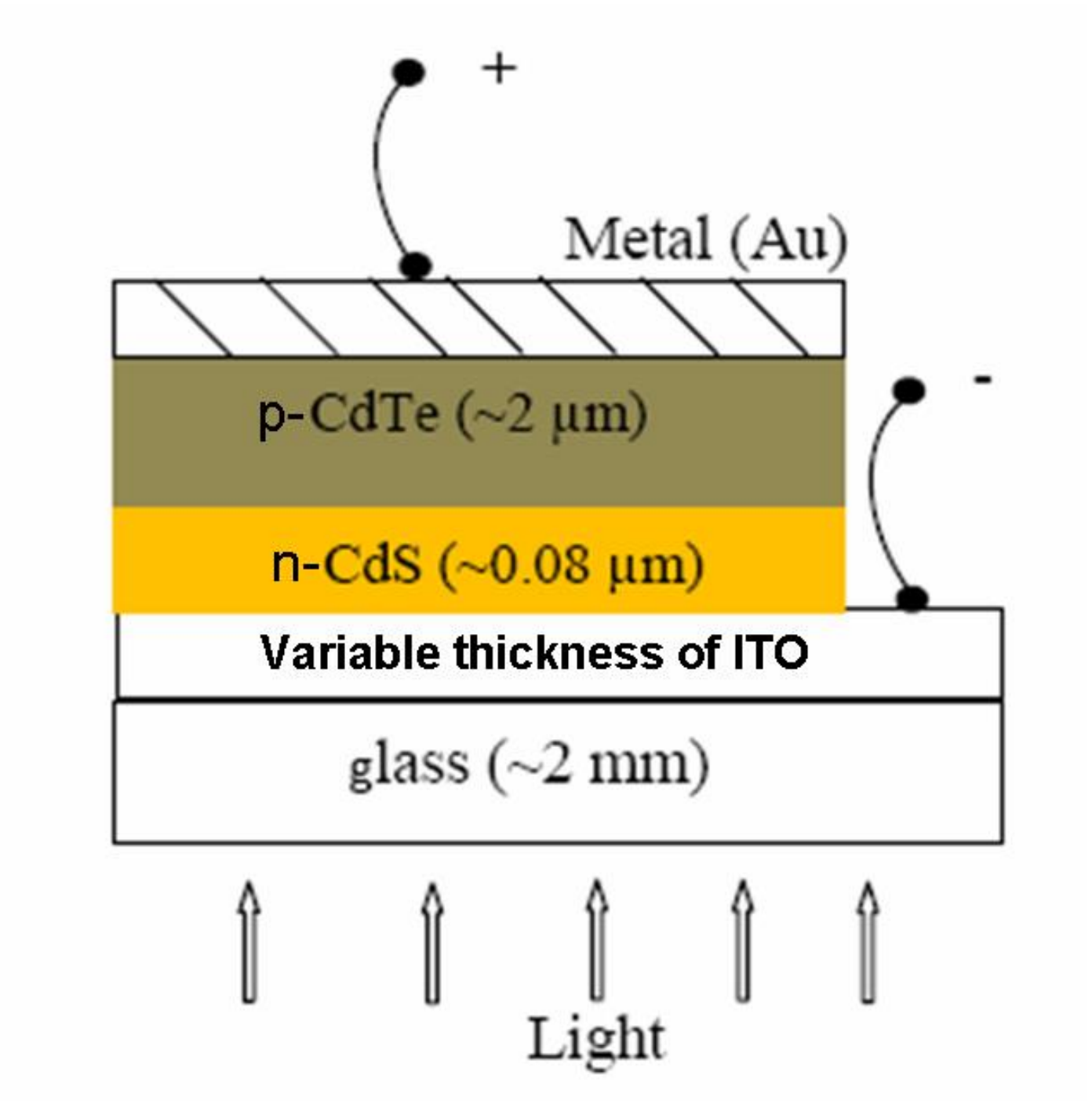

**Fig. 12:** The basic structure of the glass/ITO/CdS/CdTe/metal thin-film solar cell diagram

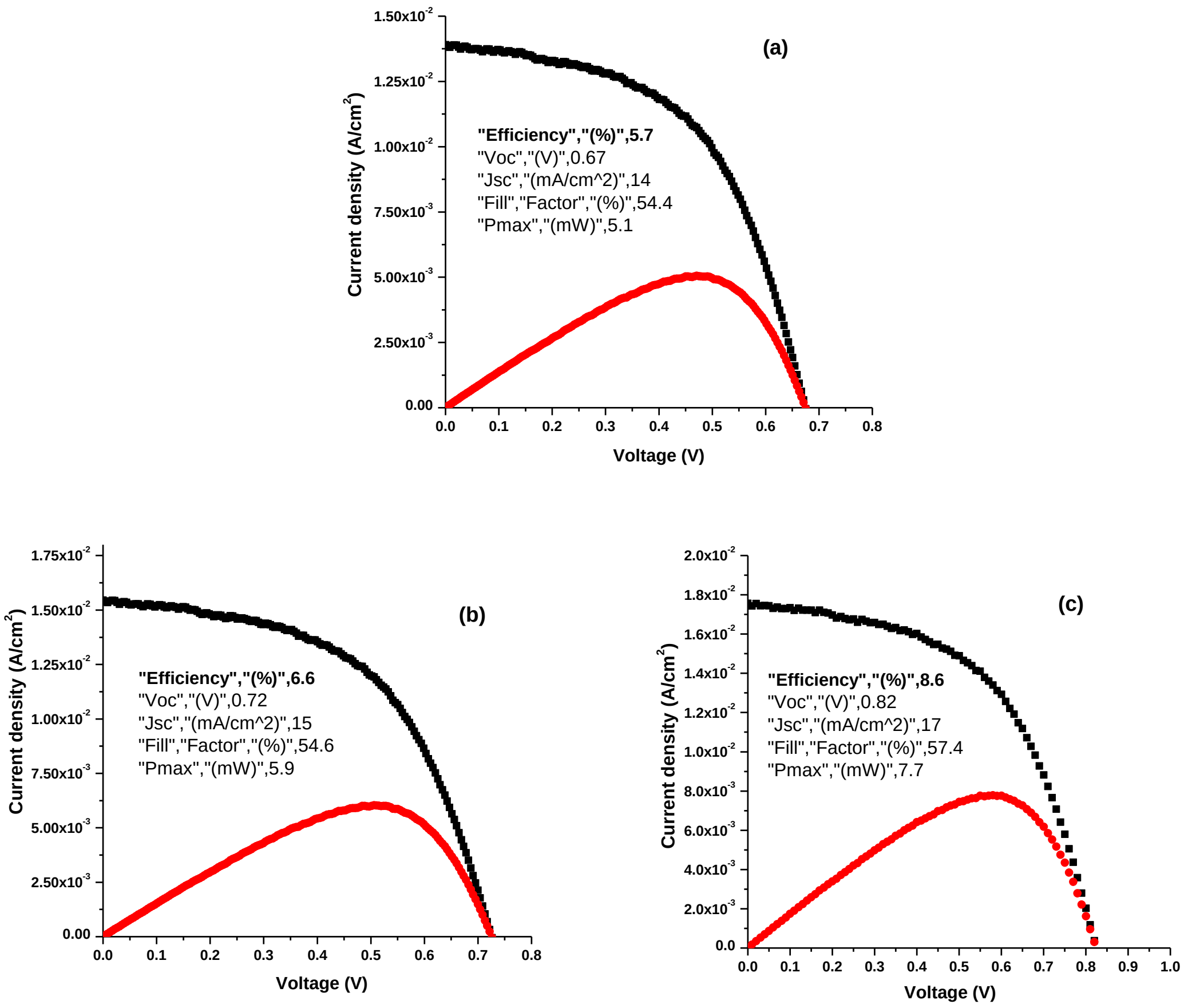

**Fig. 13:** Characteristic curves of solar cell at (a) d = 75 nm, (b) d = 225 nm and d = 325 nm.